
\magnification=\magstep 1
\baselineskip=16pt
\hsize=15.5truecm
\vsize=22.5truecm
\hoffset=1.5truecm
\font\texto=cmr10 
\font\titulo=cmbx10 scaled \magstep 1
\font\dep=cmti9


\newbox\Ancha
\def\gros#1{{\setbox\Ancha=\hbox{$#1$}%
\kern-.025em\copy\Ancha\kern-\wd\Ancha
\kern.05em\copy\Ancha\kern-\wd\Ancha
\kern-.025em\raise.0433em\box\Ancha}}

\texto

\

\

\centerline{\titulo  NOETHER'S THEOREM AND TIME-DEPENDENT}
\centerline{\titulo  QUANTUM INVARIANTS \footnote{$^\ast$}{\texto Work
supported in part by project UNAM-DGAPA IN103091. }}

\

\centerline{O. Casta\~nos, R. L\'opez-Pe\~na and V.I. Man'ko
\footnote{$^\dagger$}{\texto On leave from Lebedev Institute of
Physics, Moscow, Russia.}}
\centerline{\dep Instituto de Ciencias Nucleares, UNAM}
\centerline{\dep Apdo. Postal 70-543, 04510 M\'exico, D. F., M\'exico}

\

\

\

\noindent
{\titulo Abstract}

\

\noindent
The time dependent-integrals of motion, linear in position and
momentum operators, of a quantum system are extracted from Noether's
theorem prescription by means of special time-dependent variations of
coordinates.  For the stationary case of the generalized
two-dimensional harmonic oscillator, the time-independent integrals of
motion are shown to correspond to special Bragg-type symmetry
properties.  A detailed study for the non-stationary case of this
quantum system is presented.  The linear integrals of motion are
constructed explicitly for the case of varying mass and coupling
strength.  They are obtained also from Noether's theorem.  The
general treatment for a multi-dimensional quadratic system is indicated, and it
is shown that the time-dependent variations that
give rise to the linear invariants, as conserved quantities, satisfy
the corresponding classical homogeneous equations of motion for the
coordinates.

\

\

\

\

\noindent
{\titulo Pacs.: 02.20.+b; 42.50.Dv; 03.65.Fd}

\vfill
\eject

\noindent
{\titulo Introduction}

\vskip1.0truecm

\noindent
For some stationary and non-stationary systems the time-dependent
integrals of motion have been constructed explicitly [1-5].  On the
other hand it is well known that the integrals of the motion of
classical and quantum systems are related to the symmetry of the
system.  This relation is expressed by Noether's theorem [6,7].  The
generalized stationary two-dimensional oscillator has been analyzed
from the point of view of Noether's theorem [8].  In this work on the
base of Noether's theorem the time-independent integrals of the
motion, which are polynomials in the position and momentum operators,
were found.

Lewis and Riesenfeld [1] constructed integrals of motion quadratic in
position and momentum operators for a time-dependent
one-dimensional
oscillator and for a charged spinless particle moving in an
homogeneous varying magnetic field.  In Ref. [2] the time-dependent
integrals of motion which are linear in position and momentum
operators have been found for a charged particle moving in a
time-dependent homogeneous magnetic field.  In Ref. [3] the linear
time-dependent integrals of motion have been obtained for the quantum
forced oscillator with time-dependent frequency.  Using these
integrals of motion the dynamical symmetry $SU(2)$ and $SU(1,1)$ [3]
has been associated for the charged particle moving in a varying
magnetic field and for the parametric quantum oscillator.  In Ref.
[4] the time-dependent invariants linear in position and momentum
operators have been found for multi-dimensional quantum systems whose
hamiltonian is a non-stationary general quadratic form in position
and momentum operators.  These integrals of motion have been used to
construct the propagator, coherent states and transition amplitudes
between the energy levels of the system in terms of symplectic
transformations $ISp(2N, \hbox{I\hskip -2pt R})$ [4,5].  The
time-dependent integrals of motion give an useful method to find the
propagator of quantum systems using the system of equations found in
Ref. [9] (see also [5]).  Urrutia and Hern\'andez [10] considered a
non-stationary damped harmonic oscillator, and have proved that the
relation of linear time-dependent integrals of motion to the
propagator is close in spirit to the Schwinger action principle
method.  In Ref. [11], and more recently [12], have been emphasized
the connection of time-dependent integrals of motion of some
examples of stationary physical problems with the dynamical symmetry
concept of quantum systems.

On the other hand up to now it was not clear how the time-dependent
integrals of motion found in Ref. [1-5] can be obtained from the
canonical procedure of Noether's approach and what is the symmetry
which corresponds to linear time-dependent integrals of motion of
non-stationary multi-dimensional forced harmonic oscillator.  The aim
of the present work is to show that there exists such a variation of
coordinates for the generalized oscillator [8] as well as for the
non-stationary one-dimensional and multi-dimensional oscillators for
which lagrangian variation is reduced to a total time-derivative
term.  So from normal Noether's theorem we will obtain linear
time-dependent integrals of the motion for the generalized
two-dimensional oscillator and connect these integrals with the
analogue of Bragg scattering relation giving the time-independent
invariants found in Ref. [8].  We will consider also explicitly the
non-stationary generalized two-dimensional oscillator with varying
mass and a time-dependent coupling term proportional to the third
component of the angular momentum operator.  It is also known that
time-dependent hamiltonians [5] can generate squeezed states and
therefore we are going to study the two-mode squeezing [13-15] in the
frame of this model.  Finally we show that the analysis made for the
generalized two-dimensional oscillator can be extended for
multi-dimensional non-stationary quadratic systems.  Thus we find its
corresponding time-dependent invariants, linear in position and
momentum, by means of Noether's theorem.

\vskip 1.0truecm

\noindent
{\titulo 1. \ \ Generalized two-dimensional harmonic oscillator}

\vskip 1.0truecm

\noindent
Following Ref. [8] we will remind in this section the properties of the
integrals of motion and Noether's theorem for the two-dimensional
generalized oscillator.  We will start from the time-independent
system
$$
H = {1\over 2} \sum^2_{i=1} (p^2_i + x^2_i) + \lambda M \ ,   \eqno(1.1)
$$
where we are using dimensionless units, {\it i.e.,} $\hbar = m_0 =
\omega_0 = 1$.  This hamiltonian is constructed by a two-dimensional
harmonic oscillator plus $M = x_1p_2 - x_2 p_1$, the projection of
the angular momentum in the $O_z$ direction, with coupling constant
$\lambda$.  For $\lambda =1$ the hamiltonian (1.1) describes the
Landau problem of a charged particle moving in a constant magnetic
field.  Introducing the variables
$$
a^\dagger_i = {1\over \sqrt{2}} (x_i - ip_i) ~, ~~a_i = {1\over
\sqrt{2}} (x_i + ip_i) ~~ , ~~i = 1, 2  \ , \eqno(1.2)
$$
and
$$
\eta_\pm = {1\over \sqrt{2}} \left(a^\dagger_1 \pm
ia^\dagger_2\right) ~~, ~~\xi_\pm = {1\over \sqrt{2}} (a_1 \mp ia_2) \ ,
\eqno(1.3)
$$
which obey commutation relations
$$
[\xi_a, ~\eta_b] = \delta_{ab}; ~~ [\xi_a, \xi_b] = [\eta_a, \eta_b]
= 0 ~ , ~ a,b = +, -  \ , \eqno(1.4)
$$
we can rewrite the hamiltonian (1.1) in terms of these operators
$$
H = (1 + \lambda) \, N_+ + (1 -\lambda) \,  N_- \ , \eqno(1.5)
$$
where a constant term was neglected and $N_a~, \ a = \pm$, denotes
the number of quanta in direction $a$.  If $|\lambda|>1$ the
hamiltonian (1.1) has an energy spectrum unbounded from below.  This
spectrum, except for constant terms, is described by the formula
$$
E_{\nu m} = (1 + \lambda)n_+ + (1 - \lambda) n_- = \nu + \lambda m
\eqno(1.6)
$$
with $|m| = \nu ~, \ \nu -2 ~, \ \dots 1$ or $0$ and $\nu = n_+ +
n_-~, \ m = n_+ - n_-$, and $n_\pm = 0, 1, 2, \dots $.  The accidental
degeneracy of the hamiltonian (1.5) was explained in [8] by means of
the Noether's theorem and also the existence of time-independent
integrals of motion.  These constants of the motion, depending on the
strength of the parameter $\lambda$, are given by
$$
\eqalignno{ \eta^{k_1}_+ \eta^{k_2}_- ~, \quad \xi^{k_1}_+
\xi^{k_2}_-~, & \quad \hbox{for}\quad |\lambda| >1 \ , & (1.7a,b) \cr
\eta^{k_1}_+ \xi^{k_2}_- ~, \quad \xi^{k_1}_+
\eta^{k_2}_-~, & \quad \hbox{for}\quad |\lambda| <1 \ , & (1.7c,d) \cr
\eta_\pm \xi_\pm~, \ \quad \eta_\mp\xi_\mp ~, & \quad \hbox{for} \quad
\lambda = \mp 1 \ , & (1.7e,f) \cr}
$$
where the relative prime integers $k_1$ and $k_2$ are connected with
the rational number $\lambda$ by the formula
$$
{1 - \lambda \over 1 +\lambda} = -\epsilon {k_1 \over k_2} \eqno(1.8)
$$
with the number $\epsilon = -1$ for $|\lambda|<1$ and $\epsilon = 1$
for $|\lambda|>1$.  The question which we want to answer here is how
to find the time-dependent integrals of motion, linear in position
and momentum, for the system and how these invariants are related to
the integrals of motion (1.7).  The system with hamiltonian (1.5) is
quadratic and due to the results in  Ref. [2-5] there is a four dimensional
symplectic matrix defined by ${\bf \Lambda}$, which relates the linear
time-dependent integrals of motion with the position and momentum operators:
$$
\left(\matrix{\pi_{10}(t) \cr \pi_{20} (t) \cr  q_{10}(t) \cr
q_{20} (t) \cr} \right) = {\bf \Lambda} (t) \left( \matrix{\pi_1 \cr \pi_2
\cr q_1 \cr q_2 \cr} \right) \eqno(1.9)
$$
where we associate indices 1 and 2 with the labels $+$ and $-$.
Besides this matrix ${\bf \Lambda (t)}$ satisfies the first order differential
equation
$$
{d {\bf\Lambda}  \over dt} = {\bf \Lambda}{\bf \Sigma B} ~ , \eqno(1.10)
$$
with the initial condition ${\bf \Lambda} (0) = {\bf I}_4$.  The ${\bf
I}_4$ denotes the $4 \times 4$ indentity matrix and the matrix
$ { \bf \Sigma} $ has the form
$$
{\bf \Sigma} = \biggl( \matrix{\hfill 0 & \hfill {\bf I}_2 \cr
\hfill -{\bf I}_2 & \hfill 0 \cr}\biggr) \ , \eqno(1.11)
$$
where ${\bf I}_2$ denotes de $2\times 2$ identity matrix.  The ${\bf B}$ is
a matrix determined by the hamiltonian (1.5), and, in the case considered, it
is
$$
{\bf B} = \left( \matrix{ 1 + \lambda & 0 & 0 & 0 \cr
0 & 1 - \lambda & 0 & 0 \cr
0 & 0 & 1+\lambda & 0 \cr
0 & 0 & 0 & 1-\lambda \cr } \right) \ . \eqno(1.12)
$$
The equation (1.10) may be easily integrated and the integrals of the
motion which are linear forms in position and momentum may be written
down in the form of creation and annihilation operators
$$
A_1 (t) = \hbox{exp} \{i (1 + \lambda) \, t\} ~ A_1(0)~, \qquad A_2 (t)
= \hbox{exp}\{i (1  - \lambda) \, t\} ~ A_2(0) \ , \eqno(1.13)
$$
where we choose the initial conditions
$$
A_1(0) = \xi_+~, \qquad A_2(0) = \xi_- ~. \eqno(1.14)
$$
Then, the invariants (1.7) are immediately obtained from the
integrals of the motion (1.13) by calculating the operators
$$
\eqalignno{ A^{k_1}_1(t)A^{k_2}_2(t) & \quad \hbox{for} ~ |\lambda|>1~,
 & (1.15a) \cr
A^{k_1}_1 (t) A^{\dagger k_2}_2 (t) & \quad \hbox{for} ~|\lambda |<1~,
& (1.15b) \cr
A_1(t) = \xi_+ & \quad \hbox{for} ~\lambda = -1~, & (1.15c) \cr
A_2(t) = \xi_- & \quad \hbox{for} ~\lambda = 1~. & (1.15d) \cr}
$$
By substituting (1.13) and (1.14) into the relations (1.15a,b) and
demanding that these integrals of motion do not depend on time, we
obtain the conditions
$$
\eqalignno{ k_1(1+\lambda) + k_2(1-\lambda) = 0 ~, & \quad \hbox{for}
{}~ |\lambda | > 1 ~, & (1.16a) \cr
k_1 (1+\lambda) - k_2 (1-\lambda) = 0 ~, & \quad \hbox{for}
{}~|\lambda| < 1~. & (1.16b) \cr }
$$
These conditions remind the Bragg relation in X-ray crystallography.
Thus for some integers $k_1$ and $k_2$ the dependence on time of the
integrals of motion (1.15) disappears, in agreement with the
accidental degeneracy implied by the Eq. (1.8).

\vskip 1.0truecm

\noindent
{\titulo 2. \ Linear invariants and Noether's theorem}

\vskip 1.0truecm

\noindent
We have found in the previous section the linear invariants (1.13)
and (1.14) simply guessing their form and proving that they are
integrals of motion by direct checking.  Now we will discuss how they
can follow from Noether's theorem.  Before doing that, to illustrate
the procedure we consider first how the linear invariants for a
one-dimensional parametric oscillator found in Ref. [2,3] follow from
Noether's theorem.  The lagrangian of this oscillator has the form
$$
L = {\dot q^2\over 2} - {\omega^2(t)q^2 \over 2} ~. \eqno(2.1)
$$
Following Noether's theorem procedure used in Ref. [8] let us
consider the variation in coordinate
$$
\delta q = h (t) ~. \eqno(2.2)
$$
We have used a specific variation depending only on an arbitrary
time-dependent function $h(t)$, and it is straightforward to
calculate the induced variation in the lagrangian
$$
 \delta L  = \dot q \ (\delta q)^. - (\omega^2 q)\delta
q
= \dot q \, \dot h(t) - \omega^2 q h(t) \ . \eqno(2.3)
$$
This variation can be written as a total time derivative of a
function, $\Omega = \dot h q$, with respect to time if the function
$h(t)$ satisfies the equation
$$
\ddot h + \omega^2 (t) h = 0 \ , \eqno(2.4)
$$
which is identical to the equation of motion of the physical system.
The integrals of motion, due to Noether's theorem procedure, are
determined by the function $\Omega$ in the form
$$
K(t)  = {\partial L \over \partial \dot q} \ \delta q -
\Omega
= h \dot q - \dot hq \ . \eqno(2.5)
$$
Considering trapped particle problems, Glauber [16] pointed out that
the invariant of Ref. [2,3] may be obtained by comparing it with the
structure of a wronskian.  Since $\dot q = p$, the expression (2.5)
gives linear time-dependent integrals of the motion.  Therefore we
have obtained this invariants from Noether's theorem.

Let us now get back to the two-dimensional generalized oscillator
(1.1) as another and more complicated example.  The lagrangian for
this system has the form
$$
L(x_1~,~x_2~,~\dot x_1~,~\dot x_2) = {1\over 2} m (\dot x^2_1 + \dot
x^2_2) - {1\over 2} m (\omega^2_0 - \lambda^2) (x^2_1 + x^2_2) +
\lambda m (x_2\dot x_1 - x_1 \dot x_2) \eqno(2.6)
$$
where for later convenience we take  into account explicitly the
mass and the frequency of the oscillator, however if we put $m =
\omega_0 = 1$ we recover the case discussed in Sec. 1.  On the basis
of experience with the one-dimensional problem let us consider
a variation of coordinates $x_1$ and $x_2$ of the form
$$
\delta x_1 = h_1(t) ~, \qquad \delta x_2 = h_2(t) ~. \eqno(2.7)
$$
The corresponding induced variation in the lagrangian (2.6) is
expressed as
$$
\eqalignno{ \delta L & = m (\dot h_1 + \lambda h_2) \dot x_1 + m
(\dot h_2 - \lambda h_1) \dot x_2 \cr
& - m [\lambda \dot h_2 + (\omega^2_0 - \lambda^2) h_1] x_1 + m
[\lambda \dot h_1 - (\omega^2_0 - \lambda^2) h_2]x_2 ~. \cr}
$$
It may be rewritten in the following manner:
$$
\eqalign{ \delta L = & {d\over dt} \{ m (\dot h_1 + \lambda h_2) x_1
+ m (\dot h_2 - \lambda h_1) x_2 \} \cr
& - m \left[\ddot h_1 + {\dot m \over m} \dot h_1 + 2 \lambda \dot h_2 +
(\omega^2_0 - \lambda^2) h_1 + {(\lambda m) ^. \over m} h_2 \right]
x_1 \cr
& - m \left[\ddot h_2 + {\dot m \over m} \dot h_2 - 2 \lambda \dot h_1 +
(\omega^2_0 - \lambda^2) h_2 - {(\lambda m) ^. \over m} h_1 \right]
x_2 ~. \cr } \eqno(2.8)
$$
The equations of motion for the generalized oscillator with
lagrangian (2.6) are
$$
\eqalign{ \ddot x_1 & = (\lambda^2 - \omega^2_0) x_1 - 2\lambda \dot
x_2 - {\dot m \over m}\dot x_1 - {(\lambda m)^ . \over m} x_2 ~, \cr
\ddot x_2 & = (\lambda^2 - \omega^2_0) x_2 + 2\lambda \dot
x_1 - {\dot m \over m}\dot x_2 + {(\lambda m)^ . \over m} x_1 ~. \cr}
\eqno(2.9)
$$
Comparing the last two terms of (2.8) with (2.9), we conclude again
that if the variations of coordinates $h_1(t)$ and $h_2(t)$ satisfy
the system of equations of motion of the physical system the
variation of the lagrangian (2.6) takes the form of a full derivative
$\delta L = d\Omega/dt$, with the function $\Omega$ defined by
$$
\Omega = m (\dot h_1 + \lambda h_2) x_1 + m (\dot h_2 - \lambda h_1)
x_2 ~. \eqno(2.10)
$$
Due to Noether's theorem prescription we have the integrals of motion
of the system
$$
I(t) = {\partial L \over \partial \dot x_1} h_1 + {\partial L \over
\partial \dot x_2} h_2 - \Omega ~.
$$
These invariants are linear in positions and momenta
$$
I(t) = (p_1 + \lambda mx_2) h_1 - m \dot h_1 x_1 + (p_2 - \lambda
mx_1) h_2 - m\dot h_2 x_2 ~. \eqno(2.11)
$$
Thus we explained from Noether's theorem the existence of linear
time-dependent integrals of the motion.  There are four invariants
because there exist four independent solutions for the system of
equations (2.9).  These different solutions are denoted by a
superindex: $h^{(k)}_1, \ h^{(k)}_2$, with $1 \leq k \leq 4$.  This
set of constants of motion can be rewritten in the form (1.9), with the matrix
${\bf\Lambda}$ defined by the row vectors
$$
{\bf\Lambda}^{\left( k \right)} = \left(  h^{(k)}_1 , \, h^{(k)}_2 ,  \,
-m\left(\dot
h^{(k)}_1 + \lambda h^{(k)}_2 \right) , \,  m\left(\lambda h^{(k)}_1 -
\dot h^{(k)}_2 \right)  \right)  \ . \eqno(2.12)
$$
The constants of the motion, which we denote by $p_{i 0}$ and $x_{i 0}$,
satisfy the initial conditions $p_{10}(0) = p_1, \, p_{20}(0) =
p_2, \, x_{10} (0) = x_1$ and $x_{20} (0) = x_2$. These imply that the matrix
${\bf\Lambda} (0) = {\bf I}_4$, {\it i.e.},
	$$\eqalignno{ {\bf h}_{1} (0) & = (1, 0, 0, 0)^T \ ,
		\quad \dot{\bf h}_{1} (0) = (0, - \lambda (0), -{1
		\over m (0)}, 0)^T \ , & (2.13a) \cr
		{\bf h}_{2} (0) & = (0, 1, 0, 0)^T \ ,
		\quad \dot{\bf h}_{2} (0) = ( \lambda (0), 0, 0, - {1
		\over m (0)})^T \ , & (2.13a) \cr} $$
where $ {\bf h}_{\alpha} ( t ) = ( h^{(1)}_{\alpha} , \, h^{(2)}_{\alpha} , \,
h^{(3)}_{\alpha} , \, h^{(4)}_{\alpha} )^{T} , \ \alpha = 1 , \, 2 $.
To find the explicit form of the variations $h^{(k)}_1 (t)$ and
$h^{(k)}_2(t)$, we need to solve the classical equations of motion of
the physical system.  Let us introduce the change of variables
$$
\eqalign{z & = h_1 + ih_2 ~, \cr
z^\ast & = h_1 - ih_2 ~, \cr}\eqno(2.14)
$$
which allows to rewrite the system of differential equations for
$h_1$ and $h_2$ as
$$
\ddot z + \biggl( {\dot m \over m} - 2i\lambda \biggr) \dot z +
\biggl[ (\omega^2_0 - \lambda^2) - {i\over m} (\lambda m)^.\biggr] z
= 0~. \eqno(2.15)
$$
By means of the transformation
$$
z(t) = m ^{-1/2} \hbox{exp}\biggl\{ i \int^t_0 \ d\tau \lambda (\tau)
\biggr\} w (t) ~, \eqno(2.16)
$$
the equation (2.15) can be simplified to the form
$$
\ddot w + \Omega^2 w = 0 ~ , \eqno(2.17)
$$
with
$$
\Omega^2 = \omega^2_0 + \biggl( {\dot m \over 2m} \biggr)^2 - {\ddot
m \over 2m} ~. \eqno(2.18)
$$
The Eq.(2.17), for appropriate choices of the time dependent function
$\Omega$, can be solved and then from expressions (2.14) and (2.16)
we have the general solution for (2.9):
$$
\eqalign{ & h_1 = {1\over 2\sqrt{m}} \biggl( w \ \hbox{exp}\biggl\{ i
\int^t_0 d\tau\lambda (\tau) \biggr\} + w^\ast \ \hbox{exp} \biggl\{
- i \int^t_0 d\tau(\lambda (\tau) \biggr\}\biggr) ~ , \cr
& h_2 = {1\over 2i\sqrt{m}} \biggl( w \ \hbox{exp}\biggl\{ i
\int^t_0 d\tau\lambda (\tau) \biggr\} - w^\ast \ \hbox{exp} \biggl\{
- i \int^t_0 d\tau(\lambda (\tau) \biggr\}\biggr) ~ . \cr}\eqno(2.19)
$$

To find the linear time-dependent integrals of the motion of the
generalized harmonic oscillator with constant parameters, we take $m
= \omega_0 =1$ and $\lambda$ an arbitrary constant.  The solution of
equation (2.17) is
$$
w(t) = A \ \hbox{exp} \ (it) + B \ \hbox{exp} \ (-it) ~.
$$
{}From (2.19) we have the solutions for $h_1$ and $h_2$:
$$
\eqalignno{& h _1 (t) = |A| \cos (t + \lambda \, t + \phi_A) + | B| \cos
(t - \lambda \, t - \phi_B) ~, & (2.20a) \cr
& h _2 (t) = |A| \sin (t + \lambda \, t + \phi_A) + | B| \sin
(t - \lambda \, t - \phi_B) ~, & (2.20b) \cr }
$$
where we denoted the complex numbers $A$ and $B$ in polar form $C = |
C| \hbox{exp} (i\Phi_C)$.  Taking into account the initial conditions
given in (2.13) we arrive to the four independent solutions
$$ \eqalignno{{\bf h}_{1} {(t)} = & ( \cos\lambda t \
\cos t ~, - \sin
\lambda t \ \cos t ~, -\cos\lambda t \ \sin t ~, \sin \lambda t \ \sin t
{}~)^T \ , & (2.21a) \cr {\bf h}_{2} {(t)} = &
( \sin\lambda t \  \cos t ~, \cos \lambda t \
\cos t ~, -\sin \lambda t \ \sin t ~, - \cos \lambda t \ \sin t ~)^T \ ,
			 & (2.21b) \cr }$$
Substituting these results in (2.12), we find that the matrix ${\bf\Lambda}$
can be written in blocks of $2\times 2$ matrices $\lambda_k, \ 1\leq k \leq 4$
such that
$$
{\bf\Lambda} = \pmatrix{ \lambda_1 & \lambda_2 \cr
\lambda_3 & \lambda_4\cr} \ ,  \qquad \lambda_k = \  \mu_k \  {\bf R} ~,
\eqno(2.22a,b)
$$
where $\mu_1 = \mu_4 = \cos t$ and $\mu_2 = - \mu_3 = \sin t$.
The ${\bf R}$ is a rotation matrix by an angle $\theta = \lambda t$ , {\it
i.e.},
$$
{\bf R} = \pmatrix{ \ \ \cos \theta & \sin \theta \cr
-\sin \theta & \cos \theta \cr} ~. \eqno(2.23)
$$
Through these expressions it is immediate to get the constants of the
motion, which can be written down in the form of annihilation
operators $a_{k0} = {1\over \sqrt{2}} (x_{k0} + i p_{k0})$:
$$
\eqalignno{ & a_{10}(t) = \hbox{exp} (it) \{ a_1 \ \cos(\lambda t) +
a_2 \ \sin(\lambda t)\} ~, & (2.24a) \cr
& a_{20} (t) = \hbox{exp} (it) \{ - a_1 \ \sin(\lambda t) + a_2 \cos
(\lambda t) \} ~, & (2.24b) \cr}
$$
and their correspondent hermitean conjugates.  From them we built
 the annihilation operators (1.3) and found agreement with the
results (1.13) of the first section.

\vskip 1.0truecm

\noindent
{\titulo 3. \ Examples with Time-Dependent  Parameters}

\vskip 1.0truecm

\noindent
We study systems described by the hamiltonian
$$
H = {1\over 2} \sum_i \biggl( {p^2_i \over m} + m\omega^2_0 x^2_i
\biggr) + \lambda (x_1 p_2 - x_2 p_1) ~. \eqno(3.1)
$$
Through Noether's theorem we get the linear time-dependent integrals
of motion, and from these invariants, following Ref. [2-5], we can
evaluate the evolution operator, the associated coherent states,
and the correlation matrices in $pq$ and $a^\dagger a$ spaces.  In
these calculations for quadratic hamiltonians, the fundamental
quantity is the symplectic matrix  ${\bf \Lambda}$, which
relates linear time-dependent integrals of motion with the
momentum and position operators. Next we are going to consider the hamiltonian
(3.1) for several
choices of the parameters $m$ and $\lambda$, and determine the
analytic expressions for the $2 \times 2$ submatrices $\lambda_k$ of ${\bf
\Lambda}$.

First, we consider an exponentially varying mass $m =  m_0 \
\hbox{exp} \ \{\gamma t\}$, and $\lambda$ an arbitrary function of
time.  These parameters imply through (2.18) that $\Omega^2 =
\omega^2_0 - \gamma^2/4$, so considering $\gamma^2 < 4\omega^2_0$, we
solve Eq.(2.17), and using (2.19) together with the initial
conditions (2.13), we determine that the matrix $\gros\Lambda$ of the
system has the structure given in (2.22) but in this case the rotation angle is
$\theta = \int^t_0
\lambda (\tau) \ d\tau$ and the $\mu_k$ functions take the form
$$
\eqalignno{ & \mu_1 = \hbox{exp} \ \{-\gamma t/2\} \ \biggl(
\cos\Omega t + {\gamma \over 2} \ {\sin\Omega t \over \Omega} \biggr)
 ~, & (3.2a) \cr
& \mu_2 = \hbox{exp} \ \{ \gamma t/2\} m_0 \omega^2_0 {\sin\Omega
t \over \Omega}  ~ , & (3.2b) \cr
& \mu_3 = - {1\over m_0} \ \hbox{exp} \ \{ - \gamma t/2\} \ {\sin
\Omega t \over \Omega} ~, & (3.2c) \cr
& \mu_4 = \hbox{exp} \ \{\gamma t/2\} \ \biggl(
\cos\Omega t - {\gamma \over 2} \ {\sin\Omega t \over \Omega} \biggr)
 ~. & (3.2d) \cr }
$$
The case $\gamma^2> 4\omega^2_0$ is obtained from expressions (3.2)
by replacing
$$
\Omega \longrightarrow i\Omega_1
$$
with $\Omega_1 = \sqrt{\gamma^2/4 - \omega^2_0}$, and by means of the
elementary relations $\cos i\Omega_1 = \cosh \Omega_1$ and $\sin
i\Omega_1/i\Omega_1 = \sinh\Omega_1/\Omega_1 $.  Finally, when
$\gamma = \pm 2\omega_0$, the corresponding $\lambda_k$ matrices are
obtained from (3.2) by taking the limit when $\Omega \to 0$.

It is important to  remark that if we take $\gamma = 0$, the
$\mu_k$ functions take the form
$$
\eqalignno{ & \mu_1 = \cos \omega_0 t \ , & (3.3a) \cr
& \mu_2 = m_0 \omega_0 \sin\omega_0 t \, & (3.3b) \cr
& \mu_3 = - {1 \over m_0 \omega_0} \sin \omega_0 t \ , &
(3.3c) \cr
& \mu_4 = \cos \omega_0 t \ , & (3.3d) \cr }
$$
from which, if we take $m_0 = \omega_0 = 1 $, and $\lambda$ a constant, we
recover the result given in the previous section.

Sometimes it is convenient to write the quantum invariants in
the form of creation and annihilation operators, because the
eigenfunctions of the integrals of motion $\vec A (t) = (A_1(t), \
A_2(t))$ define solutions of the time-dependent Schroedinger equation
or coherent-type states of the system:
$$
\pmatrix{\vec A(t) \cr \vec A^\dagger (t) \cr} = {\bf M} \pmatrix{\vec a
\cr \vec a^\dagger} ~. \eqno(3.4)
$$
The matrix ${\bf M}$ is defined through the expression ${\bf M} = g
\Lambda g^{-1}$, with the $g$ matrix given by
$$
g  = {1\over \sqrt{2}} \ \pmatrix{ \ \ i {l\over \hbar} {\bf I}_2 &
{1\over l} {\bf I_2} \cr -i {l \over \hbar} {\bf I}_2 & {1\over l}
{\bf I}_2\cr} ~, \eqno(3.5)
$$
and $l = \sqrt{\hbar/_{\textstyle{ m_0 \omega_0}}}$ is the oscillator
length.  It is important to emphasize that ${\bf M}$ is also a
 symplectic matrix.  It is sometimes useful to write the matrix {\bf
M} in terms of the $2 \times 2$ matrices ${\bf M}_k, \ 1 \leq k \leq
4$:
$$
\eqalignno{ & {\bf M}_1 = {1\over 2} \biggl( \lambda_1 - im_0\omega_0
\lambda_3  + \lambda_4 + {i\over m_0 \omega_0} \lambda_2 \biggr) ~, &
(3.6a) \cr
& {\bf M}_2 = {1\over 2} \biggl(- \lambda_1 + im_0\omega_0
\lambda_3  + \lambda_4 + {i\over m_0 \omega_0} \lambda_2 \biggr) ~ , &
(3.6b) \cr }
$$
with ${\bf M}_3 ={\bf M}^{\ast}_2$ and ${\bf M}_4 = {\bf M}^{\ast}_1$.
If we consider $m = 1$, {\it i.e.}, $\gamma =0$, and $m_0 = 1$, the
operators (3.4) can be written
$$
\eqalignno{ & A_1 (t) = \hbox{exp} \{it\} \ \{a_1
\cos(\overline{\lambda} t) + a_2 \sin(\overline{\lambda}t)\} ~, &
(3.7a) \cr
& A_2 (t) = \hbox{exp} \{it\} \ \{-a_1
\sin(\overline{\lambda} t) + a_2 \cos(\overline{\lambda}t)\} ~, &
(3.7b) \cr }
$$
where $\overline{\lambda} = \int^t_0 \lambda (\tau) \
d\tau/_{\textstyle{t}}$, is the average of $\lambda$ during the
period of time $t$.  Taking linear combinations as in (1.3) we arrive
to the linear time-dependent invariants
$$
\eqalignno{ & \xi_ + (t) = \hbox{exp}\{ i(\omega_0 +
\overline{\lambda}) t \} \xi_+ ~, & (3.8a) \cr
& \xi_- (t) = \hbox{exp}\{ i(\omega_0 -
\overline{\lambda}) t \} \xi_- ~. & (3.8b) \cr}
$$
And proceeding as in (1.15) we obtain the Bragg-like conditions
$$
\eqalignno{ k_1(\omega_0 + \overline{\lambda}) + k_2 (\omega_0 -
\overline{\lambda}) = 0 ~, & \quad \hbox{for} \ |\overline{\lambda} |
> 1 ~, & (3.9a) \cr
k_1(\omega_0 + \overline{\lambda}) - k_2 (\omega_0 -
\overline{\lambda}) = 0 ~, & \quad \hbox{for} \ |\overline{\lambda} |
< 1 ~. & (3.9b) \cr }
$$

Substituting (2.22b) into the expressions for the correlation matrices
(A.6), given in the Appendix A, and using that ${\bf R}$ is an
orthogonal matrix, we have immediately that the dispersion matrices $
\sigma^2 $ are independent on the parameter $\lambda$, and are
diagonal:
$$
\eqalignno{ & \sigma^2_{pp} = {\hbar m_0\omega_0 \over 2} \biggl\{
{1\over (m_0 \omega_0)^2} \mu^2_2 + \mu^2_4\biggr\} {\bf I}_2 ~, &
(3.10a) \cr
& \sigma^2 _{pq} = - {\hbar\over 2} \biggl\{ {1\over m_0 \omega_0}
\mu_2 \mu_1 + (m_0\omega_0) \mu_4\mu_3 \biggr\} {\bf I}_2 ~, &
(3.10b) \cr
& \sigma^2_{qq} = {\hbar \over 2m_0 \omega_0} \{ \mu^2_1 + (m_0
\omega_0)^2\mu^2_3 \} {\bf I}_2 ~ . & (3.10c) \cr }
$$
Finally taking into account that for the considered system the $\mu_k$'s
functions are given by (3.2) we get

$$
\eqalignno{
\sigma^2_{pp}(t) & = {\hbar m_0\omega_0 \over 2} \ \hbox{exp} (\gamma
t) \biggl[ 1+{\gamma^2 \over 2\Omega^2} \sin^2 \Omega t - {\gamma
\over 2\Omega} \sin 2\Omega t\biggr] \ {\bf I}_2 ~, & (3.11a) \cr
\sigma^2_{qq}(t) & = {\hbar \over m_0\omega_0 } \ \hbox{exp} (-\gamma
t) \biggl[ 1+{\gamma^2 \over 2\Omega^2} \sin^2 \Omega t + {\gamma
\over 2\Omega} \sin 2\Omega t\biggr] \ {\bf I}_2 ~, & (3.11b) \cr
\sigma^2_{pq}(t) & = -{\gamma\hbar \omega_0 \over 2\Omega^2}
 \sin^2 \Omega t \ {\bf I}_2 ~. & (3.11c) \cr }
$$
For the values of the parameters $\gamma = 0.1$ and $\omega_0$ equal
to $1, \ 1/20$ and $1/30$, the behavior of the dispersion matrices is
illustrated in the Figs. 1, 2, and 3, respectively.  In these
figures, we observe that there is squeezing for the coordinates and
stretching for the momenta.  Also it is interesting to note that
$\sigma_{pq}$ is a negative function; because it is not identical to
zero, there is correlation between the coordinates and the momenta.
If we reverse the sign of $\gamma$, the roles between the dispersion
for coordinates and momenta are interchanged, and $\sigma_{pq}$
becomes positive.  In particular, for the case $\gamma^2<
4\omega^2_0$ shown in Fig. 1, the dispersion matrix $\sigma_{pq}$ is
a negative oscillating function.  Finally, the change of the mass as
a function of time, in arbitrary units, is displayed in the Fig. 4.

Now we consider a varying mass of the form
$$
m(t) = \cases{ m_0 \hfill ~~, & $t\leq 0$ ~; \cr
m_0 \cosh^2\Omega_0 t \hfill ~~, & $0\leq t\leq T$ ~; \cr
m_0 \{ \Omega_0 (t - T) \sinh \Omega_0 T + \cosh \Omega_0 T\}^2 ~~, &
$T\leq t$ ~; \cr}\eqno(3.12)
$$
which is substituted into (2.18).  We solve the ordinary differential
equation (2.17) for the indicated time ranges, and obtain
$$
w(t) = \cases{ A \ \hbox{exp} \ (i\omega_0 t) + B \ \hbox{exp} \
(-i\omega_0 t) ~~, &
$t\leq 0$\hfill ~;  \cr
C \ \hbox{exp} \ (i\tilde \Omega t) + D \ \hbox{exp} \ (-i\tilde
\Omega t)  \hfill ~~,
& $0\leq t\leq T$ ~; \cr
F \ \hbox{exp} \ (i\omega_0 t) + G \ \hbox{exp} \  (-i\omega_0 t)
\hfill ~~, & $T\leq t$~; \cr}
$$
where $\tilde\Omega \equiv \sqrt{\omega^2_0 - \Omega^2_0}$.  By
asking continuity conditions for the function $w(t)$ and its
derivative in $t = 0$ and $t = T$, we find the following relations
between the constants:
$$
\eqalignno{ & C = { A + B \over 2} + {A-B \over 2\tilde\Omega} ~ ,
\quad D = {A+B \over 2} - {A -B \over 2\tilde\Omega} ~, \cr
& F = \hbox{exp} ( - i\omega_0 T) \biggl[ {A+B\over 2} \biggl(
\cos\tilde\Omega T + i{\tilde\Omega \over \omega_0} \sin \tilde\Omega
T \biggr) + {A - B \over 2}\biggl( \cos\tilde\Omega T +
i{\tilde\omega_0 \over \Omega} \sin \tilde \Omega T \biggr) \biggr]~,
\cr
& G = \hbox{exp} (i\omega_0 T) \biggl[{A + B\over 2} \biggl( \cos
\tilde\Omega T - i {\tilde\Omega \over \omega_0} \sin \tilde\Omega
T\biggr) - {A - B \over 2} \biggl( \cos\tilde\Omega T - i
{\tilde\omega_0 \over \Omega} \sin \tilde \Omega T \biggr)\biggr] ~.
\cr }
$$
Substituting the constants $C, \ D, \ F$, and $G$, into the relation
for $w (t)$, it is straightforward but lengthy to build the general solutions
(2.19)
for (2.9).  The constants  A and B are determined through the initial
conditions (2.13).

This new system will have a ${\bf \Lambda}$ matrix, which has three
different functional forms according to the time intervals indicated in
Eq. (3.12), however in all the cases the $\lambda_k$ matrices have the
structure indicated in Eq. (2.22), with an ${\bf R}$ matrix identical to the
one of the previous example : i) The functions $\mu_k$ for $t\leq 0$ are
obtained by
taking $\gamma=0$ into the expressions (3.3), and similarly the
corresponding dispersion matrix is obtained through the equation
(3.11).  ii) For $0 \leq t\leq T$, we have to put the corresponding
solutions (2.19) into the general relation  (2.12).  The resulting
 $\mu_k$ functions are
$$
\eqalignno{ & \mu_1 = {\cos \tilde\Omega t \over \cosh \Omega_0
t}, & (3.13a) \cr
& \mu_2 = m_0 \{ \tilde\Omega \cosh \Omega_0 t \sin \tilde\Omega
t + \Omega _0 \cos \tilde \Omega t \sinh \Omega_0 t\}  , &
(3.13b) \cr
& \mu_3 =  - { \sin \tilde \Omega t \over m_0 \tilde \Omega \cosh
\Omega_0 t}  , & (3.13c) \cr
& \mu_4 = \biggl\{ \cosh \Omega_0 t \cos \tilde\Omega t - {\Omega
_0 \over \tilde\Omega} \sin \tilde\Omega t \sinh \Omega_0 t \biggr\}
. & (3.13d) \cr}
$$
Substituting these into the expressions (3.10) we arrive to the
correlation matrices for $0 \leq t \leq T$ associated to the system
(3.1), with the mass parameter of the form  (3.12):
$$
\eqalignno{ \sigma^2_{pp} & = {\hbar m_0\omega_0 \over 4} \biggl\{
{\tilde\Omega^2 \over \omega^2_0} - {\Omega^2_0 \over \tilde\Omega^2}
\sin^2 (\tilde\Omega t) - {\Omega^3_0 \over \tilde\Omega\omega^2_0}
\sinh (2\Omega_0 t) \sin(2\tilde\Omega t) \cr
& \qquad + \cosh (2\Omega_0 t) \biggl[ \sin^2(\tilde\Omega t) +
{\omega^2_0 \over \tilde\Omega^2} \cos^2(\tilde\Omega t)
\biggr]\biggr\} {\bf I}_2 ~, & (3.14a) \cr
\sigma^2_{qq} & = {\hbar \over 2m_0 \omega_0} \hbox{sech}^2 (\Omega_0 t)
\biggl[ {\omega^2_0 \over \tilde\Omega^2 } - {\Omega^2_0 \over \tilde
\Omega^2} \cos^2 (\tilde\Omega t) \biggr] {\bf I}_2 ~ , & (3.14b) \cr
\sigma^2_{pq} & = {\hbar\over 2} \biggl\{ {\Omega^2_0 \over 2\omega_0
\tilde \Omega} \sin(2\tilde\Omega t) - \biggl[ {\omega_0 \Omega_0
\over \tilde\Omega^2} - {\Omega^3_0 \over \tilde \Omega^2\omega_0}
\cos^2 (\tilde\Omega t)\biggr]\tanh (\Omega_0 t) \biggr\} {\bf I}_2
{}~. & (3.14c) \cr }
$$
For $t\geq T$ the $\mu_k$ functions are given by the following cumbersome
expressions:
$$
\eqalignno{
\mu_1 & = {1\over \sqrt{m/_{\textstyle m_0}}} \biggl\{
\cos\tilde\Omega T \cos[\omega_0 (t - T)] - {\tilde\Omega \over
\omega_0} \sin \tilde\Omega T \sin[\omega_0 (t - T)] \biggr\} ~, &
(3.15a) \cr
\mu_2 & = {\Omega_0 m_0 \over 2} \sinh (\Omega_0 T) \biggl\{ \biggl[
1 - {\tilde\Omega \over \omega_0} \biggr] \cos \left( - \omega_0 (t -
T) + \tilde\Omega T\right) \cr
& \qquad + \biggl[ 1 + {\tilde \Omega \over \omega_0} \biggr] \cos
\left( \omega_0 (t - T) + \tilde\Omega T \right) \biggr\} \cr
& \qquad + {m_0 \omega_0 \over 2} \{ \cosh (\sinh (\Omega_0 T)\}
\biggl\{ \biggl[ 1 + {\tilde \Omega \over \omega_0}\biggr] \sin
\left( \omega_0 (t - T) + \tilde\Omega T\right) \cr
& \qquad - \biggl[ 1 - {\tilde \Omega \over \omega_0} \biggr] \sin
\left( - \omega_0 (t - T) + \tilde\Omega T \right) \biggr\} ~ , &
(3.15b) \cr
\mu_3 & = - {1\over m_0  \omega_0} {1\over \sqrt{ m/m_0}} \left\{
{\omega_0 \over \tilde\Omega} \sin \tilde\Omega T \cos[\omega_0 (t -
T) ] + \cos\tilde\Omega T \sin [\omega_0 (t - T) ] \right\} ~, &
(3.15c) \cr
\mu_4 & = {\Omega_0 \over 2\tilde\Omega} \sinh (\Omega_0 T) \biggl\{
- \biggl[ 1 + {\tilde\Omega \over \omega_0}\biggr] \sin \left(
\omega_0 (t - T) + \tilde\Omega T \right) \cr
&  \qquad + \biggl[ 1 - {\tilde\Omega \over \omega_0} \biggr] \sin
\left( \omega_0 (t - T) - \tilde\Omega T\right) \biggr\} \cr
& \qquad + {\omega_0 \over 2\tilde\Omega} \{\cosh (\Omega_0 T) +
\Omega_0 (t - T) \sinh (\Omega_0 T)\} \biggl\{ \biggl[ 1 +
{\tilde\Omega \over \omega_0}\biggr] \cos \left( \omega_0 (t - T) +
\tilde\Omega T\right) \cr
& \qquad - \biggl[ 1 - {\tilde\Omega \over \omega_0} \biggr] \cos
\left( \omega_0 (t - T) - \tilde\Omega T\right) \biggr\} ~. & (3.15d)
\cr}
$$
Substituting these results into the equations (3.10) we arrive to the
correlation matrices for $t \geq T$.

To illustrate this case we make a specific choice of the parameters
for the mass.  We consider $m_0 =\omega_0 =1, \ \Omega_0 = 0.15$ and $T =
10$, which is displayed in Fig. 4.  The behavior of the dispersion
matrices for this selection of parameters
are shown in Fig. 5.  In spite of the change of the mass with respect
to time is different that in the previous example ({\it cf}.
Fig. 4), the general trends for the correlation matrices are similar.
 For examples, the $\sigma_{pp}$ is an increasing function of time
starting from its minimum value at $t\leq 0$, and there is squeezing
for the $\sigma_{qq}$.  The main difference appears in the
correlation $\sigma_{pq}$: in this case around the axis
$\sigma_{pq} = 0$ it is an oscillating function for $t$ large enough, while in
the previous
cases is negative or zero for any time.

\vskip 1.0truecm

\noindent
{\titulo 4. \ Coherent and Fock States}

\vskip 1.0truecm

\noindent
Now we are going to build a general expression for the coherent-like
states of the studied examples.  This is carried out by solving the
differential equations
$$
\vec A(t) \Phi _0 (\vec q, t) = 0 ~, \eqno(4.1)
$$
where
$$
\vec A(t) = \lambda_p \vec p + \lambda_q \vec q ~. \eqno(4.2a)
$$
The $\lambda_p$ and $\lambda_q$ are given in terms of the $\lambda_k$
matrices by
$$
\eqalignno{ \lambda_p & = {1 \over \sqrt{2\hbar m_0\omega_0}} (i
\lambda_1 + m_0 \omega_0 \lambda_3) ~, & (4.2b) \cr
\lambda_q & = \sqrt{m_0\omega_0 /2\hbar} \biggl( - {i\over
m_0\omega_0} \lambda_2 + \lambda_4\biggr) ~. & (4.2c)  \cr}
$$

The solution of (4.1) yields the vacuum state of the physical system,
which takes the form
$$
\Phi_0 (\vec q, t) = c(t) \ \hbox{exp} \ \biggr\{ - {i \over 2\hbar}
\vec q \ \lambda^{-1} _p  \ \lambda_q \  \vec q\biggr\} ~, \eqno(4.3)
$$
where $c(t)$ is the normalization constant.  The phase of the
wavefunction $\Phi_0 (\vec q, t)$ is chosen to guarantee that it
will be a solution of the time-dependent Schroedinger equation.
Afterwards some calculations, the final expression for the ground
state wavefunction is
$$
\Phi_0 (\vec q, t) = {1\over \sqrt{2\pi\hbar^2}\mu_p} \ \hbox{exp} \
\biggl\{ - {i\over 2\hbar} \ {\mu_q \over \mu_p} \vec q \cdot \vec q
\biggr\} ~. \eqno(4.4)
$$
In all the studied examples the  matrices $\lambda_k$ are of the
form (2.22), which was used in the previous expression, together with
the definition of the functions $\mu_p$ and $\mu_q$ by the relations
$$
\eqalignno{ & \mu_p = {1\over \sqrt{2\hbar m_0\omega_0}} \ (i \mu_1 +
m_0\omega_0 \mu_3) ~, & (4.5a) \cr
& \mu_q = \sqrt{m_0\omega_0/2\hbar} \ \biggl( - {i\over m_0\omega_0}
\mu_2 + \mu_4 \biggr) ~. & (4.5b) \cr}
$$

Next, we build the unitary operator
$$
\hat D(\alpha) = \hbox{exp} \left\{ \vec\alpha \cdot \vec A^\dagger -
\vec\alpha^\ast \cdot \vec A\right\} ~, \eqno(4.6)
$$
which is a power series expansion of integrals of motion and so it is
also an invariant.  Now we apply (4.6) to the vacuum wavefunction
(4.4), and after using a Baker-Campbell-Hausdorff formula [13], the
Eq.(4.1), and the action of the position and momentum operators, we
obtain the wavefunction
$$
\Phi_\alpha (\vec q, t) = \hbox{exp} \biggl\{ - {|\alpha|^2 \over 2}
+ {1\over 2} \ {\mu^\ast_p \over \mu_p} \ \vec\alpha \cdot \vec\alpha
+ {i\over \hbar\mu_p} \vec\alpha {\bf R} \vec q\biggr\} \Phi_0 ( \vec
q, t) ~. \eqno(4.7)
$$
This wavefunction is the general expression for the coherent-like
states in the coordinate representation of the discussed physical
systems.  By making the substitutions of the appropriate expressions
for the functions $\mu_p, \ \mu_q$ and $ {\bf R} $ into the Eq.(4.7), we
obtain the corresponding solutions for each case.

The wavefunction (4.7) can be expressed in terms of the
multi-dimensional Hermite polynomials [17] by making use of the
relation
$$
\hbox{exp} \ \biggl( - {1\over 2} g_2 \vec \alpha^\ast \cdot
\vec\alpha^\ast + g_3 \vec\alpha^\ast \tilde{\bf R} \vec\gamma\biggr)
= \sum^\infty_{n_1, n_2 =0} {\alpha^{\ast n_1}_1 \over n_1!}
{\alpha^{\ast n_2}_2 \over n_2!} {\bf H}^{\{g_2 {\bf I}_2\}}_{n_1,
n_2} \biggl( {g_3 \over g_2} \tilde{\bf R} \vec\gamma \biggr) ~.
\eqno(4.8)
$$
Substituting this expression into (4.7) and using the form of the
coherent-like states in the Fock-like representation, {\it i.e.},
$$
\langle n_1, n_2, t|\vec\alpha, t\rangle = {\alpha^{n_1}_1
\alpha^{n_2}_2 \over \sqrt{ n_1! n_2!}} \ \hbox{exp} \ \biggl\{ -
{|\alpha_1|^2 \over 2} - {|\alpha_2|^2 \over 2} \biggr\} ~,
\eqno(4.9)
$$
we can obtain the Fock-like states in the coordinate representation:
$$
\langle \vec q| n_1 n_2\rangle = \Phi_0 (\vec q, t) {\bf
H}^{\biggl\{-{\mu^\ast_p \over\mu_p}{\bf I}_2\biggr\}}_{n_1, n_2}
\biggl( - {i\over \hbar \mu^\ast_p} {\bf R}\vec q\biggr) ~.
\eqno(4.10)
$$
The multi-dimensional Hermite polynomial can be rewritten as a
product of two standard one-dimensional Hermite polynomials [17]:
$$
\eqalignno{ {\bf
H}^{\biggl\{-{\mu^\ast_p \over\mu_p}{\bf I}_2\biggr\}}_{n_1, n_2}
\biggl( - {i \over \hbar \mu^\ast_p} {\bf R} \vec q  \biggr) & =
\biggl( - {\mu^\ast_p \over 2\mu_p} \biggr)^{(n_1 + n_2)/2} H_{n_1}
\biggl( {1\over \sqrt{2} \hbar | \mu_p|} [\cos \theta \ q_1 + \sin
\theta \ q_2]\biggr) \cr
& \qquad \times H_{n_2} \biggl( { 1\over \sqrt{2} \hbar | \mu_p |}
[-\sin \theta \ q_1 + \cos\theta \ q_2] \biggr) ~, & (4.11) \cr }
$$
where we use the explicit expression of matrix ${\bf R}$.  These Fock
(4.10) and coherent (4.7) -like states are associated to the
integrals of the motion (4.2a) and therefore they represent squeezed
and correlated states as it was shown in the previous section.

\vskip 1.0truecm

\noindent{\titulo 5. \ General quadratic case}

\vskip 1.0truecm

\noindent
In this section we show that the time-dependent invariants, linear in
position and momentum, can be obtained through Noether's theorem
procedure.  Let us consider an arbitrary time-dependent
multidimensional forced harmonic oscillator [5]
$$
H =  {1\over 2} Q_\alpha {\cal B}_{\alpha\beta} (t) Q_\beta + {\cal
C}_\alpha Q_\alpha ~, \eqno(5.1)
$$
where we have defined the vector
$$
Q = \pmatrix{ p_1 \cr \vdots \cr p_n \cr q_1 \cr \vdots \cr q_n \cr}
{}~, \eqno(5.2)
$$
which denotes the $n$ position and the $n$ momentum operators, and
the matrices
$$
{\cal B} = \pmatrix{  A & B \cr C & D \cr} ~, \qquad {\cal C} =
\pmatrix{ F \cr G \cr}  ~, \eqno(5.3)
$$
which characterize the quadratic form for the Hamiltonian.  In Eq.
(5.3), $A, \ B, \ C, \break \ D$, stand for $n \times n$ matrices, and $F, \
G$, for $n \times 1$ matrices.  The hermiticity of the hamiltonian
implies that the  matrix ${\cal B}$ is symmetric, and this means the
following symmetry conditions over the four constituents $n\times n$
matrices:
$$
A^t = A, \quad B^t = C, \quad D^t = D  ~. \eqno(5.4)
$$
Expanding the hamiltonian (5.1) we obtain
$$
H =  {1\over 2} (A_{\alpha \beta} p_\alpha p_\beta + 2B_{\alpha\beta}
p_\alpha q_\beta + D_{\alpha\beta} q_\alpha q_\beta) + F_\alpha
p_\alpha + G_\beta q_\beta ~, \eqno(5.5)
$$
where the symmetry condition (5.4) was used.

Making a Legendre transformation and using the relation between
velocities and momenta we get the lagrangian of the system.
Following the procedure indicated in Sec. 2 to get the constants of
the motion for this system, let us propose an infinitesimal variation
of coordinates given by
$$
\delta q_\alpha = h_\alpha (t) \eqno(5.6)
$$
where ${\bf h}(t)$ is an arbitrary $n$ dimensional vector depending
on time.   The corresponding variation induced in the lagrangian of
the system can be rewritten like a total time derivative of a
function $\Omega$, if the variation of the coordinates satisfies the
differential equation
$$
\eqalign{ & \left( \dot h_\alpha A^{-1}_{\alpha\beta} \right) ^. +
\dot h_\alpha (A^{-1}B)_{\alpha\beta}  \cr
& \qquad - \left( h_\alpha (CA^{-1} )_{\alpha\beta}\right)^. -
h_\alpha (CA^{-1} B-D)_{\alpha\beta} = 0 ~,  \cr} \eqno(5.7)
$$
which represents the homogeneous classical equation of motion for the
coordinates of the
system [18].  For this symmetry transformation the associated
conserved quantities, according to Noether's theorem, are given by
$$
\eqalignno{ J & = \left( A^{-1}_{\alpha\beta} \dot q_\beta - (A^{-1}
B)_{\alpha\beta} q_\beta - A^{-1}_{\alpha\beta} F_\beta\right)
h_\alpha - A^{-1}_{\alpha\beta} \dot h_\alpha q_\beta +
(CA^{-1})_{\alpha\beta} h_\alpha q_\beta \cr
& \qquad  + \int^t \ dt
\left(A^{-1}_{\alpha\beta} \dot h_\alpha F_\beta +
(CA^{-1})_{\alpha\beta} h_\alpha F_\beta + h_\alpha G_\alpha\right)
{}~. \cr
&  & (5.8) \cr}
$$
There are $2n$ invariants because the system of equations (5.7) has
$2n$ independent solutions.  These $2n$ integrals of the motion can
be rewritten in the following matrix form
$$
\pmatrix{ {\bf p}_0 (t) \cr {\bf q}_0 (t) \cr} = \gros\Lambda (t)
\pmatrix{ {\bf p} \cr {\bf q} \cr} + \gros\Delta (t) ~, \eqno(5.9)
$$
where $\gros\Lambda$ is a symplectic matrix in $2n$ dimensions, which
is given in terms of the solutions (5.7) and the matrices
characterizing the physical system
$$
\gros\Lambda(t) = \pmatrix{ {\bf h}^{(1)} & \quad \left( {\bf
h}^{(1)} C - \dot{\bf h}^{(1)} \right) A^{-1} \cr
{\bf h}^{(2)} & \quad \left( {\bf
h}^{(2)} C - \dot{\bf h}^{(2)} \right) A^{-1} \cr
\vdots & \vdots \cr
{\bf h}^{(2n)} & \quad \left( {\bf
h}^{(2n)} C - \dot{\bf h}^{(2n)} \right) A^{-1} \cr} ~. \eqno(5.10)
$$
The time-dependent column vector $\gros\Delta$ is given by
$$
\Delta_k(t) = \int^t_0 dt \left( A^{-1}_{\alpha\beta} \dot
h^{(k)}_\alpha F_\beta + (CA^{-1})_{\alpha\beta} h^{(k)}_\alpha
F_\beta + h^{(k)}_\alpha G_\alpha\right) ~. \eqno(5.11)
$$
In expression (5.10), the superscript denotes the different solutions
for system (5.7), and these vector solutions are written
horizontally.  The initial conditions for these solutions are
$$
h^{(i)}_j(0) = \cases{ \delta^i_j \hfill ~, &  \quad
$1\leq i, j \leq n$ , \cr
0 \hfill ~, & \quad $n +1 \leq i, j \leq 2n$ , \cr} \eqno(5.12a)
$$
and for their derivatives,
$$
\dot h^{(i)}_j (0) = \cases{ C_{ij}(0) \hfill ~, & $1 \leq i, j \leq
n$ , \cr
-A_{ij}(0) \hfill ~, & $n+1 \leq i, j \leq 2n$ . \cr} \eqno(5.12b)
$$

We have proved that linear time-dependent invariants for
multi-dimensional oscillators can be obtained through Noether's
theorem.  This is achieved by considering a special variation which
represents a traslation along the classical trajectory of the
quadratic system.  To guarantee the existence of analytic solutions
for the integrals of motion, it is only necessary to solve the classical
equations of motion for this quadratic system.
Because these invariants are linear in position and momentum its
quantization is straightforward and then by means of the theory of
time-dependent invariants, the corresponding evolution operator and
other relevant quantities can be determined.

\vskip 1.0truecm

\noindent
{\titulo Conclusions}

\vskip 1.0truecm

\noindent
In this work we have found the time-dependent integrals of the
motion, linear in position and momentum, for the generalized
two-dimensional harmonic oscillator using the Noether's theorem.  We
consider in general varying mass and coupling strength in the Eq.
(1.1).  The generators of the symmetry group, for an arbitrary
time-dependent coupling strength, were obtained as a Bragg-like
condition on the linear invariants.  Using the model hamiltonian (1.1) for
specific choices of the
time-dependent parameters we got that the four-dimensional symplectic
matrix relating the invariants with the position and momentum
operators is of the form indicated in Eq. (3.10), which implies that
the correlation matrices are diagonal.  The dispersion matrices
obtained in the considered examples show squeezing and correlation.
We also have constructed the corresponding solutions of the
time-dependent Schroedinger equation.  In particular, we wrote the
coherent and Fock -like states in the coordinates representation.
Finally, we extended the procedure to find the time-dependent
integrals of the motion, linear in position and momentum, for the
multi-dimensional quadratic hamiltonian using again the Noether's
theorem.  The time-dependent variations that give rise to these
invariants satisfy the corresponding classical homogeneous equations
of motion for the coordinates.

\vskip 1.0truecm

\noindent
{\titulo Appendix A. Correlation Matrices }

\vskip 1.0truecm

\noindent
We are  going to show that by means of the matrices $\lambda_k$, it
is immediate to evaluate the $2\times 2$ correlation matrices for the
position and momentum operators [5].  Let us introduce the
four-vector notation
$$
\vec{\bf Q} = (p_1, p_2, x_1, x_2) \equiv (Q_\alpha) ~, \quad 1 \leq
\alpha \leq 4 ~. \eqno (A.1)
$$
Then the dispersion matrix for generalized coordinates $Q_\alpha$ is
$$
\sigma^2_{\alpha\beta} = {1\over 2} \langle \{ Q_\alpha, \ Q_\beta\}
\rangle - \langle Q_\alpha \rangle \langle Q_\beta\rangle ~, \eqno(A.2)
$$
where $\{ Q_\alpha,  \ Q_\beta\}$ denotes the anticommutator of
$Q_\alpha$ and $Q_\beta$.

{}From the expression for the linear invariants in terms of the matrix ${\bf
\Lambda}$, it is straightforward to arrive to
the following expression for the correlation matrices:
$$
\sigma^2_{\alpha\beta} (t) =\Lambda^{-1}_{\alpha\mu} (t)
 \Lambda^{-1}_{\beta\nu} (t) \sigma^2_{\mu\nu} (0) ~, \eqno(A.3)
$$
where ${\bf \sigma}^2(0)$ is the dispersion matrix for the initial
conditions, {\it i.e.},
$$
{\bf \sigma}^2(0) = {1\over 2} \pmatrix{ \hbar m_0 \omega_0 {\bf I}_2 & 0
\cr
0 & {\hbar \over m_0 \omega_0}{\bf I}_2\cr} ~. \eqno(A.4)
$$
Because ${\bf \Lambda}$ is a symplectic matrix, its inverse is given
by the expression ${\bf \Lambda}^{-1} = -{\bf \Sigma} {\bf \Lambda}^t {\bf
\Sigma}$
and substituting this result into the expression for
$\sigma^2_{\alpha\beta} (t)$, we get
$$
{\bf \sigma}^2(t)=  -{1\over 2} {\bf \Sigma} {\bf \Lambda}^t {\bf \Sigma} {\bf
\sigma}^2(0)
{\bf \Sigma} \gros\Lambda {\bf \Sigma} ~. \eqno(A.5)
$$
For the general quadratic case the matrix ${\bf \Lambda}$ has the form
indicated in (2.22a), and after substituting it into the
last expression we obtain the correlation matrices
$$
\eqalignno{ & \sigma^2_{pp} (t) = {1\over 2} \hbar m_0 \omega_0
\biggl( {1\over (m_0\omega_0)^2} \lambda^t_2 \lambda_2 + \lambda^t_4
\lambda_4 \biggr) ~, & (A.6a) \cr
& \sigma^2_{pq} (t) = -{ \hbar\over 2}
\biggl( {1\over m_0\omega_0} \lambda^t_2 \lambda_1 +
m_0\omega_0\lambda^t_4 \lambda_3 \biggr) ~, & (A.6b) \cr
& \sigma^2_{qq} (t) = {1\over 2} {\hbar\over m_0 \omega_0}
(  \lambda^t_1 \lambda_1 + (m_0\omega_0)^2 \lambda^t_3
\lambda_3) ~. & (A.6c) \cr }
$$
For application to quantum optics, sometimes is
convenient to express  the $2\times 2$ correlation matrices in terms of the
creation and annihilation operators, that is
$$
\eqalignno{ \sigma^2_{aa} & = {1\over 2} \biggl( {m_0\omega_0 \over
\hbar} \sigma^2_{qq} - {1\over \hbar m_0\omega_0}\sigma^2_{pp}\biggr)
+ {i\over \hbar}\sigma^2_{qp} ~, & (A.7a) \cr
\sigma^2_{aa^\dagger} & = {1\over 2} \biggl( {m_0\omega_0 \over
\hbar} \sigma^2_{qq} + {1\over \hbar m_0\omega_0}\sigma^2_{pp}\biggr)
 ~, & (A.7b) \cr
\sigma^2_{a^\dagger a^\dagger} & = {1\over 2} \biggl( {m_0\omega_0 \over
\hbar} \sigma^2_{qq} - {1\over \hbar m_0\omega_0}\sigma^2_{pp}\biggr)
+ {i\over \hbar}\sigma^2_{qp} ~. & (A.7c) \cr }
$$

\vfill
\eject

\noindent
{\titulo References }

\

\item{ [1] } R. Lewis and W. Riesenfeld. {\it J. Math. Phys. \bf 10}
(1969), 1458.

\item{ [2] } I. A. Malkin, V. I. Man'ko and D. A. Trifonov. {\it
Phys. Lett. A  \bf 30} (1969), 414.

\item{ [3] }  I. A. Malkin and V. I. Man'ko. {\it Phys. Lett. A \bf
32} (1970), 243.

\item{ [4] }  I. A. Malkin, V. I. Man'ko and D. A. Trifonov. {\it J.
Math. Phys. \bf 14}, 576 (1973).

\item{ [5] }  V. V. Dodonov and V. I. Man'ko, in: Proc. P. N. Lebedev
Physical Institute, Vol. 183.  Invariants and evolution of
non-stationary quantum systems, Ed. M. A. Markov (Nova Science,
Commack, N. Y., 1987).

\item{ [6] } L. D. Landau and E. M. Lifshitz. {\it Mechanics}, (Pergamon Press,
Oxford, 1960).

\item{ [7] }  E. Noether. {\it Nachr. K\"onig. Gesell. Wissen.
G\"ottingen, Math-Phys. Kl.} (1908), 235 (see: {\it Transport Theory
and Stat. Phys. \bf 1}  (1971), 186, for an english translation).

\item{ [8] }  O. Casta\~nos and R. L\'opez-Pe\~na. {\it J. Phys. A:
Math. Gen. \bf 25} (1992), 6685.

\item{ [9] }  V. V. Dodonov, I. A. Malkin and V. I. Man'ko. {\it
Int. Jour. Theor. Phys. \bf 14} (1975), 37.

\item{ [10] }  L. F. Urrutia and E. Hern\'andez. {\it Int. Jour.
Theor. Phys. \bf 23}, 1105 (1984).

\item{ [11] }  J. Dothan. {\it Phys. Rev. D \bf 2} (1970), 2944.

\item{ [12] }  O. Casta\~nos, A. Frank, and R. L\'opez-Pe\~na. {\it
J. Phys. A: Math. Gen. \bf 23} (1990), 5141.

\item{ [13] }  C. M. Caves and B. L. Schumaker. {\it Phys. Rev. A \bf
31} (1985), 3068; B. L.  Schumaker and C. M. Caves. {\it Phys. Rev. A
\bf 31} (1985), 3093.

\item{ [14] }  B. Yurke, S. L. McCall and J. R. Klauder. {\it Phys.
Rev. A} (1986), 4033.

\item{ [15] }  Y. S. Kim and V. I. Man'ko. {\it Phys. Lett. A \bf
157} (1991), 226.

\item{ [16] }  R. Glauber, in {\it Proc. of Quantum Optics
Conference}, Hyderabad, January 5-10, 1991, Eds. G. S. Agarwal and R.
Ingua (Plenum Press, New York,1993).

 \item{ [17] }  V. V. Dodonov, V. I. Man'ko, and V. V. Semjonov. {\it
Nuovo Cimento B \bf 83} (1984), 145.

\item{ [18] }  O. Casta\~nos, R. L\'opez-Pe\~na, and V. I. Man'ko:
{\it Linear Integrals of Motion and Noether's Theorem for
Multi-dimensional Quadratic Systems.} Submitted for publication.

\vfill
\eject

\noindent
{\titulo Figure Captions}

\

\

\noindent
{\bf Figure 1.}  Dispersion matrices behavior in coordinates and
momenta space given by Eq. (3.11), for the case $\gamma = 0.1$, and
$m_0 = 1 = \omega_0$. \hfill\break
{\bf Figure 2.}  Dispersion matrices behavior in coordinates and
momenta space given by Eq. (3.11) in the limit when $\gamma^2 \to
4\omega^2_0$.  The parameters used are $\gamma = 0.1$ and $m_0 = 1$.
\hfill\break
{\bf Figure 3.}  Dispersion matrices behavior in coordinates and
momenta space given by the analytic continuation of Eq. (3.11) when
$\gamma^2 > 4\omega^2_0$.  The parameters are $\gamma = 0.1, \ m_0 =
1$, and $\omega_0 = 1/_{\textstyle 30}$. \hfill\break
{\bf Figure 4.} Time dependence of the mass. The dashed line
corresponds to $m(t) = m_0 \ \hbox{exp} \ (\gamma t)$, and the solid
line to Eq. (3.12).  The parameters used to display the plot are $m_0
= 1, \ \gamma = 0.1, \ \Omega_0 = 0.15$, and $T=10$. \hfill\break
{\bf Figure 5.}  Dispersion matrices behavior in coordinates and
momenta space for the case (3.12), with parameters $\Omega_0 = 0.15, \
T=10$, and $m_0 = 1 = \omega_0$.

\end